\documentstyle[preprint,aps,prd]{revtex}
\tightenlines
\newcommand{\be}{\begin{equation}}
\newcommand{\br}{\begin{array}}
\newcommand{\er}{\end{array}}
\newcommand{\beq}{\begin{equation}}
\newcommand{\ee}{\end{equation}}
\newcommand{\eeq}{\end{equation}}

\newcommand{\<}{\langle}

\newcommand{\G}{\Gamma}
\newcommand{\p}{\partial}

\renewcommand{\>}{\rangle}
\def\ba{\begin{eqnarray}}
\def\ea{\end{eqnarray}}

\begin{document}

\preprint{hep-th/9910059}
\title{The antisymmetric tensor propagator in $AdS$}
\author{Iosif Bena}
\address{University of California, Santa Barbara, CA 93106 \\ iosif@physics.ucsb.edu}
\date{\today}
\maketitle

\begin {abstract}

In this brief note we construct the propagator for the antisymmetric tensor in $AdS_{d+1}$. We check our result using the Poincar\'e duality between the antisymmetric tensor and the gauge boson in $AdS_5$.  This propagator was needed for a computation which turned out to be too hard. It can be used for computing various other things in $AdS$.

\end {abstract}

\section{Introduction} 

In \cite{freedman,dhok-fred}, a lot of effort was put into finding the $AdS$ propagators for the graviton and the gauge boson. Their methods can be used straightforwardly for the $B_{\mu \nu}$ propagators.
An ansatz can be made for bitensor propagators \cite{allen}. This ansatz contains both gauge artifacts and gauge invariant parts. Upon using the equation of motion for $B_{\mu\nu}$ we obtain an equation for the gauge invariant part of the propagator, whose solution is hypergeometric. For d=5 it simplifies to an algebraic function of the chordal distance. As explained in \cite{freedman}, working on the subspace of conserved sources makes gauge fixing unnecessary.
We check our result by verifying the 5-dimensional Poincar\'e duality between $A_\mu$ and $B_{\mu\nu}$.
 
\section{The $B_{\mu\nu}$ propagator}
In Euclidean $AdS_{d+1}$, with the metric
$$ds^2={1 \over z_0^2}(dz_0^2+\Sigma_{i=1}^d{dz_i^2} ), \eqno(1)$$
the easiest way to express invariant functions and tensors is in terms of the chordal distance:
$$u \equiv {(z_0-w_0)^2+(z_i-w_i)^2 \over 2 z_0 w_0}. \eqno(2)$$
The action for an antisymmetric 2-tensor coupled to a conserved source $S_{\mu \nu}$ is:
$$S_B= \int{d^{d+1}z \sqrt{g}[{1\over 2\cdot 3!}H^{\mu\nu\rho} H_{\mu\nu\rho}-{1\over 2}B_{\mu\nu} S^{\mu \nu}]}, \eqno(3)$$
where
$$ H_{\mu\nu\rho}=D_{\mu}B_{\nu\rho}+D_{\nu}B_{\rho\mu}+D_{\rho}B_{\mu\nu} . \eqno(4)  $$
The Euler Lagrange equation has a solution of the form:
$$B_{\mu\nu}(z)={1\over 2}\int{d^{d+1}w\sqrt{g}G_{\mu\nu;\mu'\nu'}(z,w) S^{\mu'\nu'}(w)}, \eqno(5)$$
where $ G_{\mu\nu;\mu'\nu'}$ is the bitensor propagator. To simplify notation, the $D$'s with unprimed indices mean covariant derivatives with respect to $z$, and those with primed indices with respect to $w$.
The equation $G_{\mu\nu;\mu'\nu'} $ satisfies is:
$$D^{\rho}(D_{\mu}G_{\nu\rho;\mu'\nu'}+ D_{\nu}G_{\rho\mu;\mu'\nu'}+ D_{\rho}G_{\mu\nu;\mu'\nu'} )= -\delta(z,w)(g_{\mu\mu'}g_{\nu\nu'}- g_{\mu\nu'}g_{\nu\mu'}) + $$
$$+ D_{\mu'}\Lambda_{\mu\nu;\nu'}- D_{\nu'}\Lambda_{\mu\nu;\mu'}, \eqno(6) $$
where  $\Lambda_{\mu\nu;\nu'}$ is a diffeomorphism whose contribution vanishes when integrated against the covariantly conserved source $S^{\mu \nu}$. We can see that all of our bitensors are antisymmetric at both points.

Similarly to the methods in \cite{freedman} we observe that a suitable basis for antisymmetric bitensors
is given by:
$$T^1_{\mu\nu;\mu'\nu'} = \p_{\mu}\p_{\mu'}u  \p_{\nu} \p_{\nu'}u -\p_{\mu} \p_{\nu'}u \p_{\nu}\p_{\mu'}u \eqno (7)$$
$$T^2_{\mu\nu;\mu'\nu'} = \p_{\mu}\p_{\mu'}u  \p_{\nu}u \p_{\nu'}u -\p_{\mu}\p_{\nu'}u \p_{\nu}u \p_{\mu'}u - \p_{\nu}\p_{\mu'}u \p_{\mu}u \p_{\nu'}u +\p_{\nu}\p_{\nu'}u \p_{\mu}u \p_{\mu'}u.$$
Thus, an ansatz for $G$ is $G=T^1F^1(u)+T^2F^2(u)$.
Nonetheless, we use a different decomposition, which illustrates better the gauge artifacts
$$G_ {\mu\nu;\mu'\nu'} = T^1_{\mu\nu;\mu'\nu'} H(u) + D_{\mu}V_{\nu;\mu'\nu'}- D_{\nu}V_{\mu;\mu'\nu'}, \eqno(8) $$
where $ V_{\mu;\mu'\nu'}= Y(u)[ \p_{\mu}\p_{\mu'}u  \p_{\nu'}u -\p_{\mu} \p_{\nu'}u \p_{\mu'}u] $. Also, an antisymmetric $\Lambda_{\mu\nu;\nu'}$ can be 
expressed as 
$$\Lambda_{\mu\nu;\nu'}=A(u)[ \p_{\nu}\p_{\nu'}u  \p_{\mu}u -\p_{\mu} \p_{\nu'}u \p_{\nu}u ]. \eqno(10)$$

We can now substitute (8) and (10) in (6), and after a long computation we obtain
$$ D^{\rho}(D_{\mu}G_{\nu\rho;\mu'\nu'}+ D_{\nu}G_{\rho\mu;\mu'\nu'}+ D_{\rho}G_{\mu\nu;\mu'\nu'} ) - D_{\mu'}\Lambda_{\mu\nu;\nu'}+D_{\nu'}\Lambda_{\mu\nu;\mu'}  = $$
$$ = T^1[H''u(u+2)+H'(1+u)(d-1)-2A]-T^2[H''(1+u)+H'(d-1)+A']. \eqno(11)  $$
For $z \neq w$, we obtain 2 equations by setting the scalar coefficients of the two tensors to 0. We can observe that the $V_{\mu;\mu'\nu'}$ part which was a gauge artifact dropped out as expected.
Thus, for $u \neq 0$ we have the equations:
$$ H''u(u+2)+H'(1+u)(d-1)-2A = 0 \eqno(12a) $$
$$ H''(1+u)+H'(d-1)+A' = 0. \eqno(12b)$$
The second equation can be integrated once, with the integration constant chosen so that $A$ and $H$ vanish as $u \rightarrow  \infty$.  Combining this with (12a) we find the differential equation obeyed by $H$:
$$u(2+u)H''(u)+(d+1)(u+1)H'(u)+ 2(d-2) H = 0. \eqno(13)$$
This equation is hypergeometric, but the solution which vanishes as $u \rightarrow \infty$ is rational:
$$H(u)={\G({(d-1)/2})\over 4 \pi^{(d+1)/2}}{u+1 \over [u(u+2)]^{(d-1)/2}}, \eqno(14)$$
properly normalized to take care of the $\delta$ function in (6).

\section{Poincar\'e duality}
In 5 dimensions a 2-form is Poincar\'e dual with a gauge boson, by the relation:
$$H_{\mu\nu\rho} \epsilon^{\mu\nu\rho\sigma\lambda}= 3 ! \ F^{\sigma\lambda}  \eqno(15)$$
Therefore, we expect:
$$\<{ F^{\sigma\lambda}(z) F^{\sigma'\lambda'}(w)}\> = {1 \over (3 !)^2}  \epsilon^{\mu\nu\rho\sigma\lambda} \epsilon^{\mu'\nu'\rho'\sigma'\lambda'}  \< {H_{\mu\nu\rho}(z) H_{\mu'\nu'\rho'}(w)} \>. \eqno(16)$$
Checking (16) is a verification that our result is true.
We use the fact that 
$$\<{B_{\mu\nu}B_{\mu'\nu'} }\> = G_{\mu\nu;\mu'\nu'} \eqno(17)$$ and 
$$\<{A_{\mu}A_{\mu'}}\> = G_{\mu;\mu'}, \eqno(18)$$
where the second propagator was found in \cite{dhok-fred}.
We could check the tensor equality (16) term by term, but it is messy. We rather observe that the right hand side of (16) is a bitensor antisymmetric at both ends, and thus it will have the structure
$$\epsilon^{\mu\nu\rho\sigma\lambda} \epsilon^{\mu'\nu'\rho'\sigma'\lambda'}  \< {H_{\mu\nu\rho}(z) H_{\mu'\nu'\rho'}(w)} \> = F_1(u) T_1^{\mu\nu;\mu'\nu'} + F_2(u) T_2^{\mu\nu;\mu'\nu'}. \eqno(19)$$
Concentrating on the components of $\<{ F^{z_0z_i}(z) F^{z_0'z_j'}(w)}\>$ we obtain: 
$$2F_1+F_2(1+u) = H'', \eqno(20a)$$
$$F_2(1+u)^2+F_2+2F_1(1+u)=2H''(1+u)+3H' ,\eqno(20b)$$
which give the same $F_1$ and $F_2$ as the ones obtained from the gauge propagator derived in \cite{dhok-fred}.

\section{Conclusion}
We computed the propagator for $B_{\mu\nu}$ in $AdS_{d+1}$ and checked our result by using Poincar\'e duality for $d=4$.  This propagator can be used for computing various quantities having to do with $B_{\mu\nu}$ charged objects (like strings or D-branes with electric flux) in $AdS$. The propagators for higher form fields can also be found by using Poincar\'e duality \cite{iosif} or by explicit calculation \cite{asad}.

{\bf Acknowledgements:} I'd  like to acknowledge useful conversations with  Joe Polchinski, Gary Horowitz and Veronika Hubeny. This work was supported in part by NSF grant PHY97-22022.

\appendix
\section{Several useful identities involving the chordal distance}

In the computations the following identities were useful:

$$ \p_{\mu} \p_{\nu'}u = {-1 \over z_0 w_0}[\delta_{\mu\nu'}+{(z-w)_\mu \delta_{\nu'0}\over w_0}+	{(w-z)_\nu' \delta_{\mu0}\over z_0} - u\delta_{\mu 0}\delta_{\nu' 0}] \eqno(A1)$$
$$ \p_{\mu} u = {1 \over z_0}[(z-w)_{\mu}/w_0-u \delta_{\mu 0}]	\eqno(A2)$$
$$ \p_{\nu'} u = {1 \over w_0}[(w-z)_{\nu'}/z_0-u \delta_{\nu' 0}] \eqno(A3)$$
$$ D^{\mu}\p_{\mu}u=(d+1)(u+1)					 \eqno(A4)$$
$$ 	\p^{\mu}u\  \p_{\mu}u = u(u+2)				\eqno(A5)$$
 $$ 	 D_{\mu}\p_{\nu}u	=g_{\mu\nu}(u+1)		\eqno(A6)$$
$$ 	(\p^{\mu}u)( D_{\mu}\p_{\nu}\p_{\nu'} u) = \p_{\nu}u \p_{\nu'} u 		\eqno(A7)$$
$$ 	(\p^{\mu}u)( \p_{\mu}\p_{\nu'}) u =( u+1) \p_{\nu'} u			\eqno(A8)$$
$$ 	 D_{\mu}\p_{\nu}\p_{\nu'} u = 		g_{\mu\nu}  \p_{\nu'} u			\eqno(A9)$$

\end {document}